\documentclass[10pt]{article}

\usepackage{amsmath}
\usepackage{amssymb}
\usepackage{graphicx}
\usepackage{cite}
\usepackage[usenames,dvipsnames]{xcolor}
\usepackage{color} 

\usepackage[latin1]{inputenc}
\usepackage{textcomp}
\usepackage{amsthm}
\usepackage[english]{babel}
\usepackage{verbatim}
\usepackage{makeidx}
\usepackage{appendix}
\usepackage{indentfirst}
\usepackage{fancyhdr}
\usepackage{subfigure}
\usepackage{dcolumn}
\usepackage{bm}
\usepackage{floatrow}

\usepackage[labelfont=bf,labelsep=period,justification=raggedright]{caption}
\makeatletter
\renewcommand{\@biblabel}[1]{\quad#1.}
\makeatother

\date{}
\newcommand{\kp}{k_+}

\begin{document}

\begin{center}

\thispagestyle{empty}

{\Large\textbf {Accurate encoding and decoding by single cells: \\ amplitude versus frequency modulation}}
\\
\vspace{2mm}
Gabriele Micali$^{1,2,3}$, 
Gerardo Aquino$^{1,2}$, 
David M. Richards$^{1,2}$, and
Robert G. Endres$^{1,2,\ast}$
\\
\vspace{2mm}
$^1$ Department of Life Sciences, Imperial College, London, UK
\\
$^2$ Centre for Integrative Systems Biology and Bioinformatics, Imperial College, London, UK
\\
$^3$ Dipartimento di Fisica, Universit\` a degli Studi di Milano, Milano, Italy
\\
$^\ast$ E-mail: r.endres@imperial.ac.uk
\end{center}

\begin{abstract} 


Cells sense external concentrations and, via biochemical signaling, respond by regulating the expression of target proteins. Both in signaling networks and gene regulation there are two main mechanisms by which the concentration can be encoded internally: amplitude modulation (AM), where the absolute concentration of an internal signaling molecule encodes the stimulus, and frequency modulation (FM), where the period between successive bursts represents the stimulus. Although both mechanisms have been observed in biological systems, the question of when it is beneficial for cells to use either AM or FM is largely unanswered. Here, we first consider a simple model for a single receptor (or ion channel), which can either signal continuously whenever a ligand is bound, or produce a burst in signaling molecule upon receptor binding. We find that bursty signaling is more accurate than continuous signaling only for sufficiently fast dynamics. This suggests that modulation based on bursts may be more common in signaling networks than in gene regulation. We then extend our model to multiple receptors, where continuous and bursty signaling are equivalent to AM and FM respectively, finding that AM is always more accurate. This implies that the reason some cells use FM is related to factors other than accuracy, such as the ability to coordinate expression of multiple genes or to implement threshold crossing mechanisms.

\end{abstract}

\renewcommand{\abstractname}{\vspace{-\baselineskip}}

\begin{abstract} 
\begin{center}
{\bf Author Summary}
\end{center}
\vspace{2.5mm}

Signals, and hence information, can generally be transmitted either by amplitude (AM) or frequency (FM) modulation, as used, for example, in the transmission of radio waves since the 1930s. Both types of modulation are known to play a role in biology with AM conventionally associated with signaling and gene expression, and FM used to reliably transmit electrical signals over large distances between neurons. Surprisingly, FM was recently also observed in gene regulation, making their roles less distinct than previously thought. Although the engineering advantages and disadvantages of AM and FM are well understood, the equivalent question in biological systems is still largely unsolved. Here, we propose a simple model of signaling by receptors (or ion channels) with subsequent gene regulation, thus implementing both AM and FM in different types of biological pathways. We then compare the accuracy in the production of target proteins. We find that FM can be more accurate than AM only for a single receptor with fast signaling, whereas AM is more accurate in slow gene regulation and with signaling by multiple receptors. Finally, we propose possible reasons that cells use FM despite the potential decrease in accuracy.

\end{abstract}


\section*{Introduction}

Cells are exposed to changing environmental conditions and need to respond to external stimuli with high accuracy, e.g. to utilize nutrients and to avoid lethal stresses \cite{Swain_msbREW,Kussell_Science}. To represent (encode) chemicals in the environment, either ligand-bound receptors trigger chemical signals or ion channels allow entry of secondary messengers. These in turn activate transcription factors (TFs), which then regulate target-protein production (decoding). In eukaryotic cells, the conventional view is that the level of signaling within the cell directly encodes the external stimuli, with consequent gradual changes in the nuclear TF concentrations. This is effectively an amplitude modulation (AM) mechanism \cite{AMeFM, BlackLeff83, Mackeigan05, YuNAT08, Bosisio06, WernerHoffmann08, Giorgetti_Tiana10, OShea}. However, recent single-cell experiments also show pulsating signals \cite{AMeFM,Paszek_10, DAndrea94, Berridge97, BerridgeGalione} and bursty entry of TFs into the nucleus \cite{AMeFM, Elowitz_08, OShea,  Burst, Levine2013}, in close analogy to frequency modulation (FM). (Note that, although there is no  modulation of an underlying carrier wave as in radio broadcasting \cite{connor1973}, the AM/FM terminology is commonly used in quantitative biology \cite{Elowitz_08,OShea}.) Although several hypotheses have been put forward, the benefits and detrimental effects of either type of response remain largely unclear.

\begin{figure}[!ht]
  \begin{center}
    \includegraphics[width=0.8\textwidth]{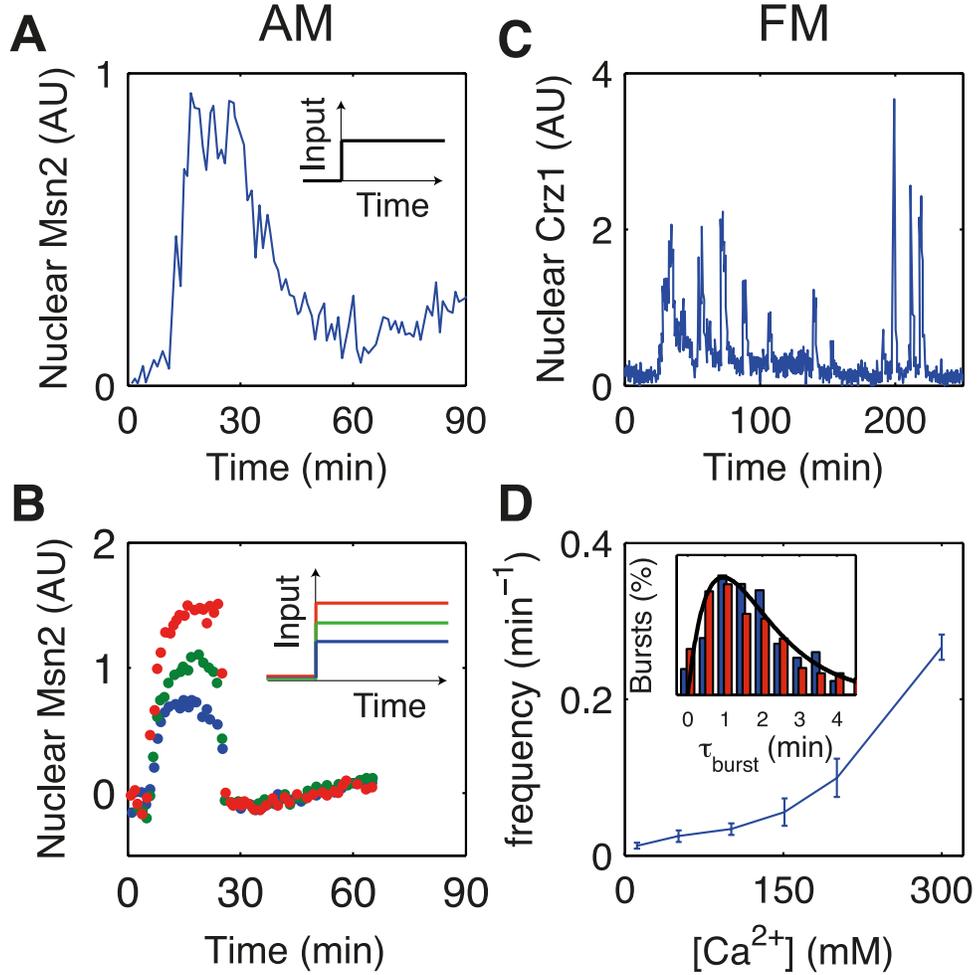}
  \end{center}
  \caption{\textbf{Experimental evidence for amplitude and frequency modulation.} (A and B) Example data showing amplitude modulation from \cite{OShea}. (A) Single-cell nuclear localization of Msn2 transcription factor in response to $\text{H}_2\text{O}_2$ stress as a function of time. The stimulus profile (input) is a step change applied at $t=0$ (inset) which applies to all figure panels. (B) Average time trace for different concentrations of $\text{H}_2\text{O}_2$ stress. (C and D) Example data showing frequency modulation from \cite{Elowitz_08}. (C) Single-cell nuclear localization of Crz1 in response to calcium stress as a function of time, showing bursts of Crz1. (D) The average frequency of bursts against calcium concentration, showing an increased frequency with increased concentration. (Inset) Burst duration distribution for low (blue bars) and high (red bars) concentration. Both histograms are well described by the Gamma distribution $h(t)=te^{-t/\tau_b}$, with $\tau_b=70$s (black solid line), demonstrating that pulse duration is independent of calcium concentration. Experimental data in arbitrary units (AU) of fluorescence.}
  \label{fig1}
\end{figure}

There is experimental evidence that both types of modulation occur in gene regulation. For example, take the budding yeast \textit{Saccharomyces cerevisiae}. Under oxidative stress the nuclear concentration of transcription factor Msn2 is proportional to the $\text{H}_2\text{O}_2$ concentration, suggesting an AM mechanism (Fig. \ref{fig1}A,B) \cite{OShea}. However, in response to a calcium stimulus, Crz1, which is normally cytoplasmic, enters the nucleus in unsynchronized bursts, regulating at least a hundred target genes (Fig. \ref{fig1}C) \cite{Elowitz_08}. The level of stimulus affects only the frequency of bursts, not their amplitude and duration, which implies FM (Fig. \ref{fig1}D and inset) \cite{Elowitz_08,Eldar_Elowitz_10}. Similarly, Msn2 and its homologue Msn4 exhibit FM under glucose limitation \cite{OShea}. Bursty FM is also found in bacteria and mammals, indicating that this is a general modulation scheme across different cell types. For example, during energy-depletion stress, the bacterium \textit{Bacillus subtilis} activates the alternative sigma factor $\sigma^{\text{B}}$ in discrete stochastic pulses, regulating around 150 downstream genes \cite{Locke}. In addition, isoform NFAT4 in activated T-cells shows similar behavior \cite{Yissachar2013}.

\begin{figure}[!ht]
  \begin{center}
    \includegraphics[width=0.8\textwidth]{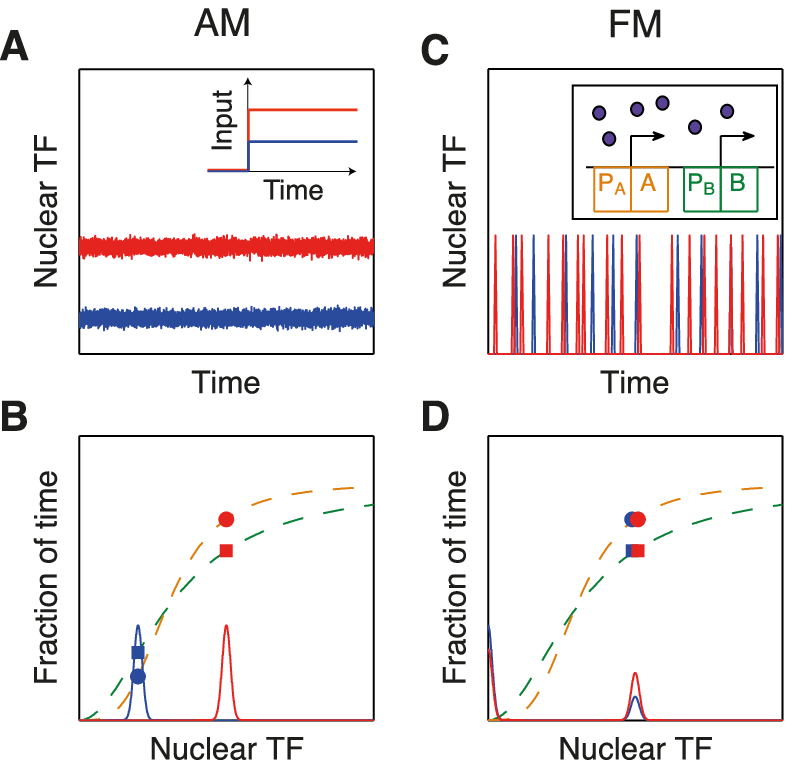}
  \end{center}
  \caption{\textbf{Advantages and disadvantages of amplitude and frequency modulation.} AM may be less noisy than FM (A,B), but FM may allow coordinated expression of many genes (C,D) \cite{Elowitz_08,Eldar_Elowitz_10}. (A) In AM, low/high stimuli result in low/high levels of transcription factor (TF) inside the nucleus. (B) In AM, different nuclear TF concentrations (blue and red curves) lead to gene expression of proteins A and B (see orange and green promoter functions respectively) with variable ratios (order of dot and square changes). (C) In FM, the stimulus strength only affects the frequency of bursts, not their amplitude. (Inset) Schematic of TF (purple dots) binding promoter $P_A$ of gene $A$ (orange) and promoter $P_B$ of gene $B$ (green) with different binding strengths. (D) In FM, the nuclear TF concentration is always the same during a burst, only the frequency of occurrence changes. As a consequence, the protein ratio stays constant.}
  \label{fig2}
\end{figure}

What are the relative benefits of AM and FM? 
One important issue is the susceptibility to noise, which affects the accuracy of sensing. For example, in broadcasting radio signals it is well known that FM is less affected by noise than AM. This is because noise mainly deteriorates the amplitude, which is where the information is stored in AM. A similar argument also favors action potentials in communicating neuronal signals over long distances \cite{Lisman_1997}. 
In contrast, it has been hypothesized that for other cell types, such as yeast, the bursty nature of FM may introduce more noise than AM, so that AM might be preferable (Fig. \ref{fig2}A,B) \cite{Elowitz_08}. However, two recent articles (which we discuss below) disagree with this and suggest that FM may still be more accurate \cite{MW,Wolde_12}. In addition, it is important to remember that there are often other factors than noise minimization. For example, it has been suggested that, in situations where multiple genes need to be up or down regulated, FM can provide greater coordination and reliability (Fig. \ref{fig2}C,D) \cite{Elowitz_08,Eldar_Elowitz_10}.

Mora and Wingreen considered a model for a single receptor embedded in a cell membrane and compared the noise in the output for two signaling mechanisms: continuous (CM) and bursty (BM) modulation \cite{MW}. In CM, the receptor signals continuously whenever a ligand is bound, whereas in BM the receptor signals for a short, fixed-sized burst only upon binding of a ligand. As we explain below, for multiple receptors these mechanisms become equivalent to AM and FM, respectively. By considering integral feedback control, a common network for sensing concentration ramps and precise adaptation \cite{Doyle_onIFC, Barkai97,Giovanna_13}, it was found that, for fast binding and unbinding, the noise in CM can be twice that from BM, suggesting that FM leads to greater accuracy. Despite this unexpected result, there are two key points that need further clarification. First, the response was only calculated to lowest order in the small-ramp parameter, thus neglecting any time dependence of the noise. Second, the derivation solely relied on the small-noise approximation, which might work well for fast signaling, but could be inadequate for slow gene regulation.

Similarly, Tostevin \textit{et al.} found biologically relevant parameter regimes of promoter switching in gene regulation in which an oscillating input can produce a more constant and hence less noisy protein output level than a constant input with noise \cite{Wolde_12}. Although interesting, this model is restricted to decoding and linear pathways, and requires fine-tuning. Its general applicability remains unclear, such as whether an oscillatory input signal can be replaced by random bursts and still remain more accurate than a constant input. In fact, oscillating signals are well-known to maximize target responsiveness while bypassing desensitization from constant signals \cite{Li_Goldbeter}. They also globally entrain with its period robust to noise \cite{Rapp_1981}. Such oscillators are found in circadian clocks, segmentation clocks, cell cycle, p53 DNA repair pathways, as well as nuclear factor NF-$\kappa$B, epidermal growth factor ERK, cAMP and $\text{Ca}^{2+}$ signaling  \cite{Tyson2011, McWatters1999, Dolmetsch_1998, Batchelor_2011, Ashall_Science, Tay2010, Albeck2013, Cai2014, Gregor_2010, Goldbeter_2012, Levine2013}. This leaves the question of the relative benefits of AM and FM (with respect to random bursts) largely unanswered.

Here, we aim to investigate the advantages and disadvantages of CM and BM (AM and FM) for encoding and decoding of constant concentrations and ramps. To build intuition, we start with a single receptor/ion channel (CM and BM). We consider concentration sensing by a linear pathway, allowing us to gain exact results for different temporal regimes (as suitable for fast signaling and slow gene regulation). 
To provide analytical results, we extend the single-receptor model for ramp sensing by Mora and Wingreen. First, we introduce an alternative mechanism to integral feedback, the incoherent feedforward loop (another common pathway motif for ramp sensing and precise adaptation \cite{Takeda_2012, Shen_Alon, Goentoro_Alon}).  This allows us to generalize the model to more than one pathway. 
Second, by explicitly including the time-dependence of signaling noise, we are able to provide first-order analytical results for the accuracy of ramp sensing. 
Taken together, a general principle emerges, favoring BM for fast signaling and CM for slow gene regulation. Finally, we generalize to many receptors and ion channels, a far more realistic situation for biological systems, allowing us to make connection with AM and FM. While we found that AM is generally more accurate than FM, we speculate why cells may still utilize FM in certain cases of gene regulation.


\section*{Results}

\begin{figure}[!ht]
  \begin{center}
    \includegraphics[width=0.8\textwidth]{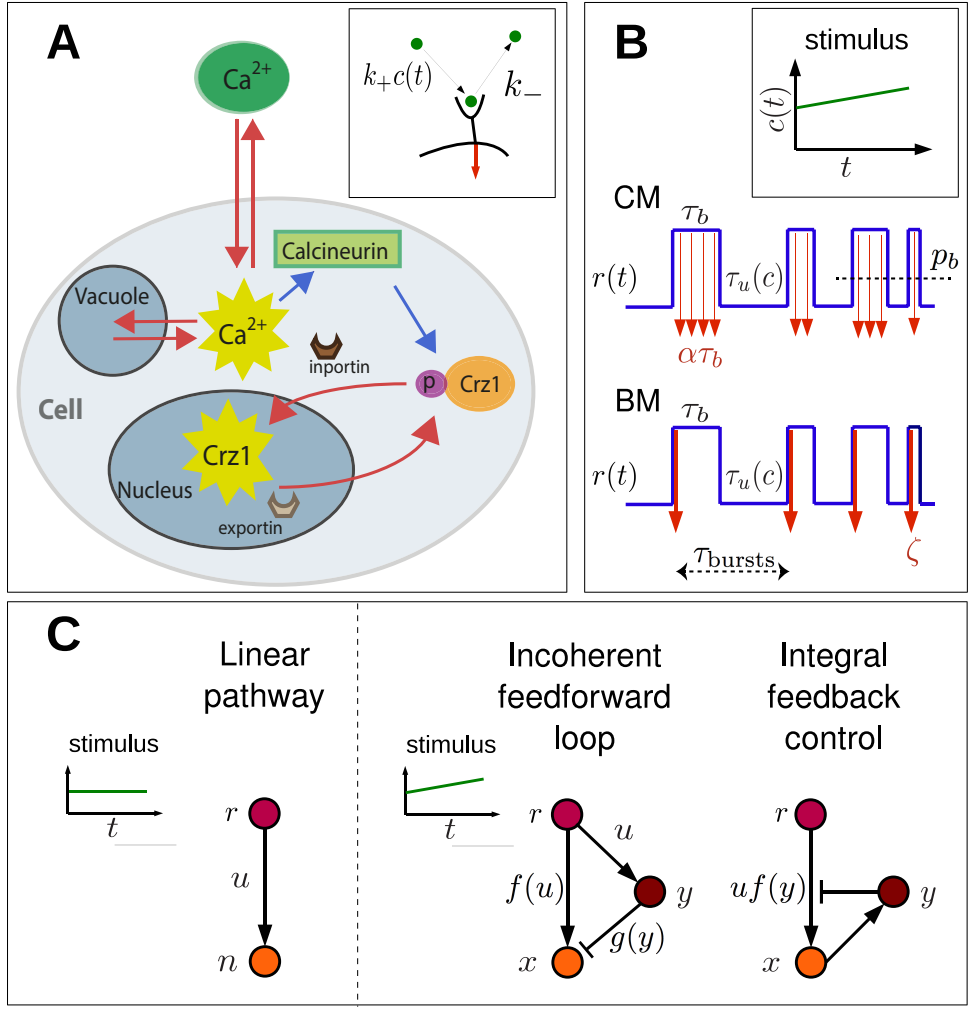}
  \end{center}
  \caption{\textbf{Schematic view of signaling and gene regulation.} (A) Cartoon of \textit{S. cerevisiae} in presence of extracellular calcium, considered a paradigm of bursty frequency modulation. Calcium enters through plasma-membrane ion channels and can be stored (released) in (from) vacuoles. Intracellular calcium activates calcineurin, which dephosphorylates Crz1p. Once dephosphorylated, Crz1 binds inporting Nmd5p and enters the nucleus. Exportin Msn5p subsequently removes Crz1 from the nucleus. Cytoplasmic calcium pulses may correspond to Crz1 bursts in the nucleus \cite{Elowitz_08}. Red arrows indicate movement while blue arrows stand for chemical signaling. (B) Single receptor/ion channel activity, $r(t)$ (blue line), depends on the concentration of extra-cellular stimulus $c$. The signaling rate $u$ differs between continuous (CM) and bursty modulation (BM). In CM, $u$ is constant rate $\alpha$ during bound intervals, with $p_b$ the probability of being bound. In BM, $\zeta$ molecules are realized at the time of binding with $\tau_{\text{bursts}}$ the duration between consecutive bursts (binding events). (C) Different regulatory networks.  Linear pathway used for concentration sensing. Incoherent feedforward loop and integral feedback control allow chemical ramps to be sensed.}
  \label{fig3}
\end{figure}

Cells sense external stimuli with cell-surface receptors and/or ion channels, which ultimately lead to changes in the concentration and dynamics of active transcription factors (TFs) inside the nucleus. Cells control the response at two different levels. Firstly, cell-surface receptors signal to regulate the activity of TFs in the cytoplasm. Secondly, inportin and exportin regulate the entry of active TFs into the nucleus, thereby regulating transcription (Fig. \ref{fig3}A). Here, we build a theoretical model that encodes information from an extra-cellular environment in an intra-cellular representation. 
We distinguish two ways of encoding this information: continuous modulation (CM) and bursty modulation (BM). Once the information is encoded, various proteins can act together to implement a response (decoding), involving regulatory networks. 
To provide a general analysis for arbitrary noise we first address concentration sensing in a simple linear pathway using the master equation.  However, to derive analytical results for ramp sensing and pathways with feedback we apply the small-noise approximation. 
We finally extend these models to implement amplitude (AM) and frequency modulation (FM) for many receptors or ion channels. Accuracy is assessed by comparing the protein output noise for the different modulation schemes, assuming that the signal is decoded by the average concentration.

\subsection*{Single-receptor/ion-channel model}

Following Mora and Wingreen \cite{MW} we build a single-receptor model that implements CM and BM. We call the extra-cellular species $c$, which is encoded intra-cellularly by the signaling rate $u$. Assuming we are in the fast diffusion regime in which each ligand molecule can bind the receptor only once, the receptor can be in either of two uncorrelated states: \textit{on} when bound and \textit{off} when unbound. This allows the receptor activity, $r(t)$, to be written mathematically as a binary response, which takes value 1 in the \textit{on} state and 0 in the \textit{off} state. The extra-cellular concentration $c$ affects the unbound time intervals $\tau_u$, such that the binding rate is given by $\left<\tau_u\right>^{-1}=k_+ c(t)$, where $k_+$ is the binding rate constant. In contrast, the bound time intervals, $\tau_b$, are exponentially distributed random numbers with average $\left<\tau_b\right>^{-1}=k_-$, where $k_-$ is the unbinding rate constant, which is independent of the extra-cellular stimulus concentration (inset in Fig. \ref{fig3}A). As for ion channels, some are ligand-gated or regulated by receptors, while others are voltage-gated and hence dependent on action potentials \cite{Hille}. In all these cases the stimulus affects the opening or closing times. In CM downstream proteins are produced with a constant rate $\alpha$ during each \textit{on} time interval, which leads to a signaling rate $u_{\text{CM}}=\alpha r(t)$, while in BM $\zeta = \alpha k_-^{-1}$ molecules are produced instantly at the moment of binding with rate $u_{\text{BM}}=\zeta \sum \delta(t-t^+_i)$, where $t^+_i$ are the binding times (Fig. \ref{fig3}B). 
This choice for $\zeta$ allows a meaningful comparison of CM and BM as both produce, on average, the same amount of intracellular species.

\subsection*{General approach to concentration sensing exhibits two regimes of accuracy}

In order to provide a general result for arbitrary input fluctuations, we write down the chemical master equation. For simplicity, we only consider concentration sensing with $c(t)=c_0$, but the model can also be applied to ramps. Furthermore, we assume a linear pathway in which the receptor/ion channel activity $r$ directly regulates an output species with copy number $n$ (with production rate $u$ and degradation rate $\gamma$) (Fig. \ref{fig3}C, left).
Since the receptor/ion channel activity is a two-state system ($\textit{on}/\textit{off}$), there are two resulting master equations for CM (one for each state) describing the probability of being in the \textit{on} and \textit{off} states, i.e. $p_{\text{on}}(n,t)$ and $p_{\text{off}}(n,t)$:
\begin{subequations}
\begin{align}
\frac{d p_{\text{on}}(n,t)}{d t} =& \ \gamma (n+1)p_{\text{on}}(n+1,t) + \alpha p_{\text{on}}(n-1,t)+ k_+ c p_{\text{off}}(n,t) - (\gamma n + \alpha + k_-)p_{\text{on}}(n,t) ,
\label{eq:masterAMon} \\
\frac{d p_{\text{off}}(n,t)}{d t} =& \ \gamma (n+1)p_{\text{off}}(n+1,t) + k_- p_{\text{on}}(n,t) - (\gamma n + k_+ c )p_{\text{off}}(n,t).
\label{eq:masterAMoff}
\end{align}  
\end{subequations}
Note that $\alpha \ge k_-$, so molecules are generally produced in the \textit{on} state.
In BM, instead, the master equations which describe the probabilities $p_\textit{on}(n,t)$ and $p_\textit{off}(n,t)$ of having $n$ proteins at time $t$, are given respectively by 
\begin{subequations}
\begin{align}
\frac{d p_{\text{on}}(n,t)}{d t} =& \ \gamma (n+1)p_{\text{on}}(n+1,t) + k_+ c p_{\text{off}}(n-\zeta,t) - (\gamma n + k_-)p_{\text{on}}(n,t) ,
\label{eq:masterFMon} \\
\frac{d p_{\text{off}}(n,t)}{d t} =& \ \gamma (n+1)p_{\text{off}}(n+1,t) + k_- p_{\text{on}}(n,t) - (\gamma n + k_+ c )p_{\text{off}}(n,t) ,
\label{eq:masterFMoff}
\end{align} 
\end{subequations}
with burst size $\zeta$ a positive integer. 
We solve Eqs. \eqref{eq:masterAMon} and \eqref{eq:masterAMoff} with generating functions and simulate Eqs. \eqref{eq:masterFMon} and \eqref{eq:masterFMoff} with the Gillespie algorithm (see Materials and Methods).

\begin{figure}[!ht]
  \begin{center}
    \includegraphics[width=0.8\textwidth]{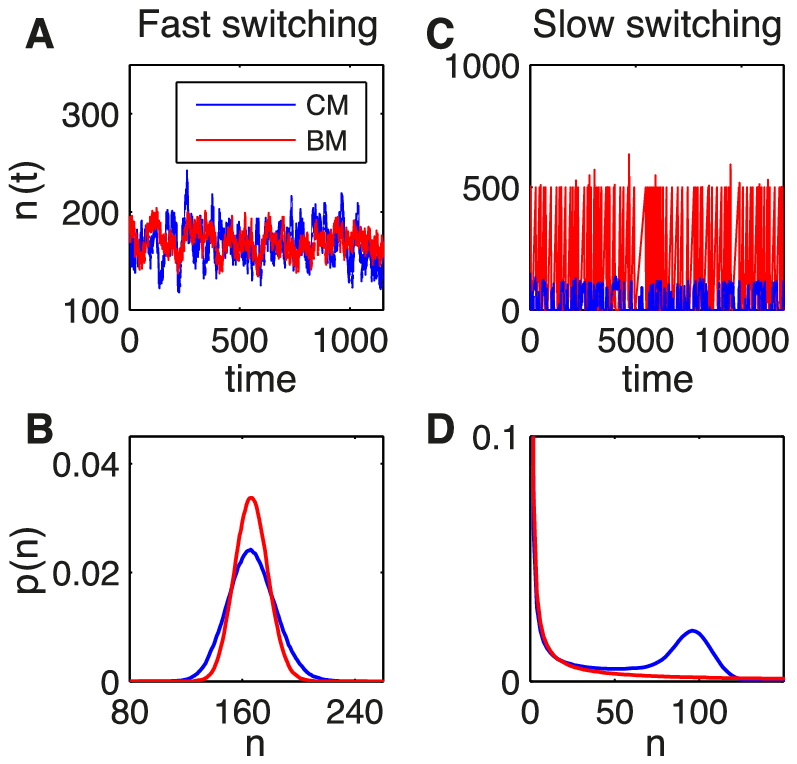}
  \end{center}
  \caption{\textbf{The two regimes in the linear pathway model based on the master equation.} (A-B) fast ($k_+ c_0=20 s^{-1}$, $k_-=100 s^{-1}$, $\gamma=0.1 s^{-1}$, $\alpha=100 s^{-1}$, $\zeta=1$) and (C-D) slow ($k_+ c_0=0.01 s^{-1}$, $k_-=0.05 s^{-1}$, $\gamma=1 s^{-1}$, $\alpha=25 s^{-1}$, $\zeta=500$) switching. (A,C) Protein number as a function of time from Gillespie simulations for CM (blue lines) and BM (red lines). (B) The probability distribution for $n$ target proteins is unimodal for both AM (blue) and FM (red). (D) The probability distribution is bimodal for AM (blue) and remains unimodal for BM (red) but with a long tail in the slow switching regime.}
  \label{fig7}
\end{figure}

Simulations via the Gillespie algorithm show different outcomes for fast (small-noise approximation limit, Fig. \ref{fig7}A,B) and slow (Fig. \ref{fig7}C,D) dynamics of the receptor. For fast switching ($k_+ c_0, k_- \gg \gamma$), for both CM and BM, the probability has an unimodal distribution (Fig.  \ref{fig7}B). On the other hand, in the slow switching regime ($k_+ c, k_- \ll \gamma$), the probability distribution becomes bimodal for CM and unimodal with a long tail for BM, leading to drastically increased noise (Fig. \ref{fig7}D). The unimodal distribution for BM, which is simply due to the use of infinitely short pulses, would become bimodal for finite width pulses. 

\begin{figure}[!ht]
  \begin{center}
    \includegraphics[width=0.8\textwidth]{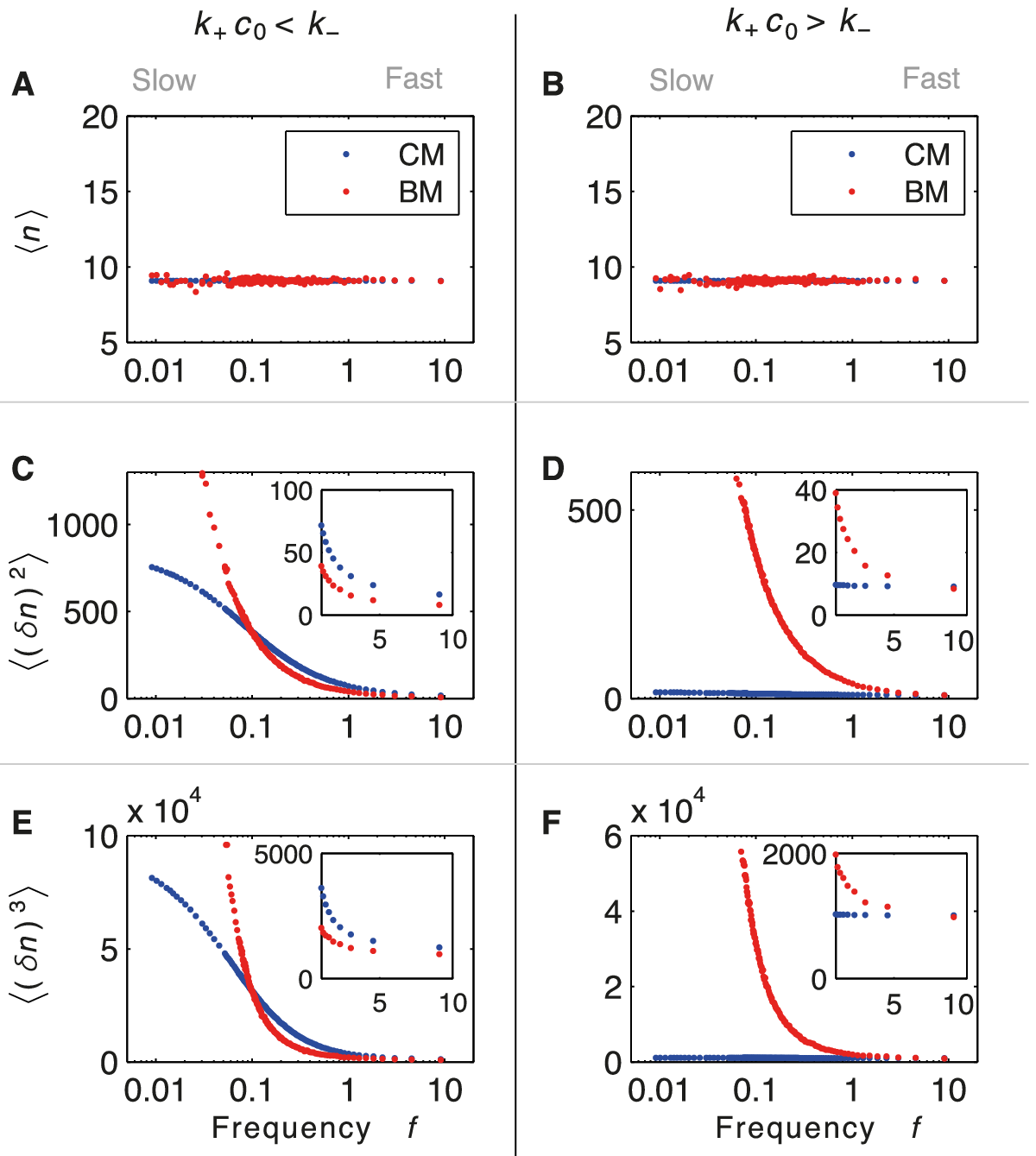} 
  \end{center}
  \caption{\textbf{First three moments of the protein distribution in concentration sensing from the master equation.} Averages (A,B), variance (C,D), and skewness (E,F) as a function of the frequency of binding events, $f=k_+ c_0/(1+k_+ c_0/k_-)$. (Insets) Magnification of small-noise approximation region (fast switching). Analytical results for CM (blue) and numerical results for BM (red) as function of the frequency of binding events (logarithmic scale). Two regimes are shown: $k_-=10 \ k_+ c_0$ ($\alpha=100 s^{-1}, \gamma=1 s^{-1}$, $\zeta \text{ from } 1000 \text{ to } 1$) (left column) and $k_-= 0.1 \ k_+ c_0$ ($\alpha=10 s^{-1}, \gamma=1 s^{-1}$, $\zeta \text{ from } 1000 \text{ to } 1$) (right column). Averages from CM and BM are constrained to be equal, \textit{i.e.} $\zeta =\alpha k_-^{-1} $. Variances of CM and BM exhibit two different regimes for fast switching: for $k_+ c_0 < k_-$ BM is more accurate than CM (inset in C), while for $k_+ c_0 > k_-$ CM is generally more accurate (inset in D), except for $\zeta=1$. Third moments show that, for large noise, the probability distributions become asymmetric.}
  \label{fig8}
\end{figure}

In order to classify the different dynamics and to compare CM and BM for arbitrary noise, we require information on the probability distribution of $n$ output proteins. In particular, we study the average, variance and skewness (the latter is encoded in the third moment) of the distribution for both CM and BM. 
Constraining the average output of CM and BM to be the same (Fig. \ref{fig8}A,B), we identify two regimes for fast dynamics: $k_+ c_0 < k_-$ (Fig. \ref{fig8}C) and $k_+ c_0 > k_-$ (Fig. \ref{fig8}D). 
Specifically, for $k_+ c_0 < k_-$, BM is more accurate (Fig. \ref{fig8}C, inset), while CM is generally more accurate when $k_+ c_0 > k_-$ (Fig. \ref{fig8}D, inset), except for minimal burst size ($\zeta=1$). However, for slow dynamics (and hence large noise), CM is always more accurate than BM. The study of the third moment shows that, for slow switching and hence bimodality, BM has large asymmetry (Fig. \ref{fig8}C,F). 

These observations can be explained as follows, using the fact that the receptor/ion channel can only detect information from the extra-cellular environment during unbound (\textit{off}) time intervals, as the extra-cellular stimulus only affects the binding rate (Fig. \ref{fig3}). For fast dynamics, the two regimes can be understood by comparison with maximum-likelihood estimation (MLE), the most accurate strategy for encoding \cite{EndWing}. MLE estimates the ligand concentration $c_{\text{ML}}=k_+^{-1} \left< \tau_u \right>^{-1}$ from the average unbound time interval $\left<\tau_u\right>$. The bound time intervals are discarded as they only contribute noise \cite{EndWing}. 
BM, which produces fixed-size bursts at the times of binding, approaches MLE when the bound intervals are shorter than the unbound intervals. In this case, the times of the bursts effectively estimate the unbound time intervals (Fig. 3B, bottom) and BM is more accurate than CM. 
However, when the bound intervals are longer than the unbound intervals, BM cannot estimate the unbound time intervals anymore and becomes less accurate than CM. Since CM produces protein during the bound intervals, it signals according to the average receptor activity $p_b=\left<\tau_b\right>/(\left< \tau_b \right> + \left< \tau_u \right>)$ (Fig. 3B, top). Hence, CM effectively contains information on both bound and unbound intervals, and thus can still provide a reasonable estimate of unbound time intervals. An interesting exception is $\alpha k_-^{-1}=\zeta=1$, for which BM becomes slightly more accurate than CM. In the latter case, since the rate of protein production during a bound interval in CM is very low, there is uncertainty as to whether CM actually produces protein or not, which reduces its accuracy. In contrast, for slow switching the burst size needs to increase since BM produces the same level of protein as CM. Hence, BM is always less accurate than CM, independent of whether bound or unbound time intervals are longer. While we analytically demonstrate the connection with MLE for fast dynamics in the next section, an extended discussion without comparison to MLE can be found in S1 Text and S1-S3 Figs.

\subsection*{Small-noise approximation to ramp sensing confirms two regimes for fast dynamics}

To further investigate fast dynamics, we extend an analytical model for ramp sensing in the small-noise approximation \cite{MW}. Considering the single-receptor described in Fig. \ref{fig3}A,B, we linearize the system by averaging over a time much larger than the binding and unbinding times. We further assume exponential distributions for $\tau_{b}$ and $\tau_{u}$ so that $\left< ( \delta \tau_b )^2 \right> = \left< \tau_b \right>^2 = k_-^{-2}$ and $\left< ( \delta \tau_u )^2 \right> = \left< \tau_u \right>^2 = (k_+ c(t) )^{-2}$, where $c(t)$ increases only very slowly with time (see below). 
Hence, signaling noise arises in CM due to variable bound time intervals (ignoring stochastic production of protein during bound intervals), while in BM the binding times (bursting times) vary.
Without loss of generality, we set $\alpha=k_-$, which is equivalent to $\zeta=1$. Hence, as we show in S1 Text, for averaging time much longer than $k_-^{-1}$ and $(k_+ c(t))^{-1}$, the average and autocorrelation (variance) of $u(t)$ are given by \cite{MW}
\begin{align}
\left< u(t) \right> =& \ \frac{k_+ c(t)}{1+ k_+ c(t) / k_-}, \label{eq:u_t} \\
\left< \delta u(t) \delta u(t') \right> =& \ g \frac{k_+ c(t)}{\left( 1+ k_+ c(t) / k_- \right)^3}\,\delta(\ t-t'), \label{eq:du_tdu_t}
\end{align} 
with $\left< \delta u(t) \right> = 0$ and
\begin{equation} \label{eq:g}
g=
\begin{cases}
1+\left< (\delta \tau_b)^2 \right> / \left< \tau_b \right>^2  &= 2  \ \ \ \ \ \ \ \ \ \ \ \ \ \ \ \ \ \ \ \ \ \ \ \ \ \text{CM} \\
1+\left< (\delta \tau_b)^2 \right> / \left< (\delta \tau_u)^2 \right>  &=1+ \left[ k_+ c(t) / k_- \right]^2  \ \ \ \ \  \text{BM}. 
\end{cases} 
\end{equation}
Note that only the variance differs between CM and BM. In particular, in Eq. \eqref{eq:g} the ratio $k_+ c(t) / k_-$ determines whether $g$ is larger in BM or CM, which ultimately determines which scheme leads to the least noise. BM has the lower noise only when $k_+ c(t) < k_-$, \textit{i.e.} when $\left< \tau_b \right> < \left< \tau_u \right>$. In particular, in the limit of fast unbinding ($k_+ c(t) \ll k_-$), the signaling noise for CM is twice as large as for BM.

Sensing temporal ramps, \textit{i.e.} the change of concentration with time, is crucial for locating nutrients and avoiding toxins. We start by considering a stimulus whose concentration is constant for $t<0$ and increases linearly and slowly in time after $t=0$:
\begin{equation} \label{eq:c}
c(t)=
\begin{cases}
c_0 & t<0 \\
c_0+c_1 t & t \ge 0,
\end{cases}
\end{equation}
for constants $c_0$ and $c_1$ with $ c_1 t \ll c_0$. By applying Eq. \eqref{eq:c} to Eqs. (\ref{eq:u_t}-\ref{eq:g}), the signaling rate can be rewritten to first order as
\begin{equation}
\label{eq:u}
u(t) \simeq
\begin{cases}
u_0+u_1 t+\delta u & t \ge 0 \\
u_0+\delta u & t<0,
\end{cases}
\end{equation}
where $u_0$, $u_1$ are functions of $c_0$ and $c_1$, and $\delta u$ is the noise described by $\left<\delta u(t)\delta u(t')\right>$ (given in S1 Text). The condition $c_1 t \ll c_0$ is necessary so that $u$ behaves linearly in time with $u_1 t \ll u_0$. Under this condition, the factor $g_\text{BM}$ of Eq. \eqref{eq:g} becomes
\begin{equation} \label{eq:defg*} 
g_\text{BM} \simeq \underbrace{1+ \frac{k_+^2 c_0^2}{k_-^2}}_{g^*_\text{BM}} + \frac{2 k_+^2 c_0 c_1}{k_-^2} t,
\end{equation}
where $g_\text{BM}$ is given by $g^*_\text{BM}$ for a constant external concentration. 
We now assume that the extra-cellular stimulus is encoded in the signaling rate $u$ which affects the production of two output proteins with concentrations $x$ and $y$. Specifically, we compare the output noise of $x$ and $y$ between CM and BM using the incoherent feedforward (Fig. \ref{fig3}C, middle) and integral feedback (Fig. \ref{fig3}C, right) loops.
\\

\subsubsection*{Incoherent feedforward loop}

The incoherent feedforward loop is a network motif in which $u$ directly affects two outputs $x$ and $y$, while $y$ inhibits $x$ (Fig. \ref{fig3}C, middle). The loop provides precise adaptation to a step-change in stimulus and can also be used for ramp sensing. Mathematically, we use the following two coupled stochastic differential equations,
\begin{align}
\label{eq:incoherentFLx}
\frac{d x}{dt} &= k_x \left( \frac{f(u)}{g(y)} - x \right), \\ 
\label{eq:incoherentFLy}
\frac{d y}{dt} &=  u - k_y y ,
\end{align}
where $k_x$  is the rate constant for production and degradation of $x$, while $k_y$ is the rate constant for degradation of $y$, and $f(u)$ and $g(y)$ are specified functions. In order to have adaptation the variable $y$ needs to evolve slower than $x$, which requires $k_x>k_y$. Here we choose $f(u)=e^{b u}$ and $g(y)=e^{b k_y y}$, where constant $b$ has units of time. This allows us to obtain an analytic solution (see S1 Text for details).

\subsubsection*{Integral feedback loop}

The integral feedback loop \cite{MW} is another network motif for precise adaptation and ramp sensing. Here, $u$ affects $x$ only (the main output), while $x$ activates $y$ and $y$ inhibits $x$ (Fig. \ref{fig3}C, right). The general equations for this model are given by
\begin{align}
  \label{eq:integralFCx}
  \frac{d x}{dt} &= u f(y) - k_x x , \\
  \label{eq:integralFCy}
  \frac{d y}{dt} &= k_y \left( x -1 \right) ,
\end{align}
where $k_{x}$ is the rate constant for degradation of $x$, $k_{y}$ is the rate constant for production and degradation of $y$ satisfying $k_x>k_y$, and $f(y)$ is a monotonically decreasing function of $y$. Specifically, we choose $f(y)=e^{-b y}$, where $b$ is a dimensionless constant. This again produces an analytic solution (see S1 Text for details).

\subsubsection*{Small-noise approximation}

\begin{figure}[!ht]
  \begin{center}
    \includegraphics[width=0.8\textwidth]{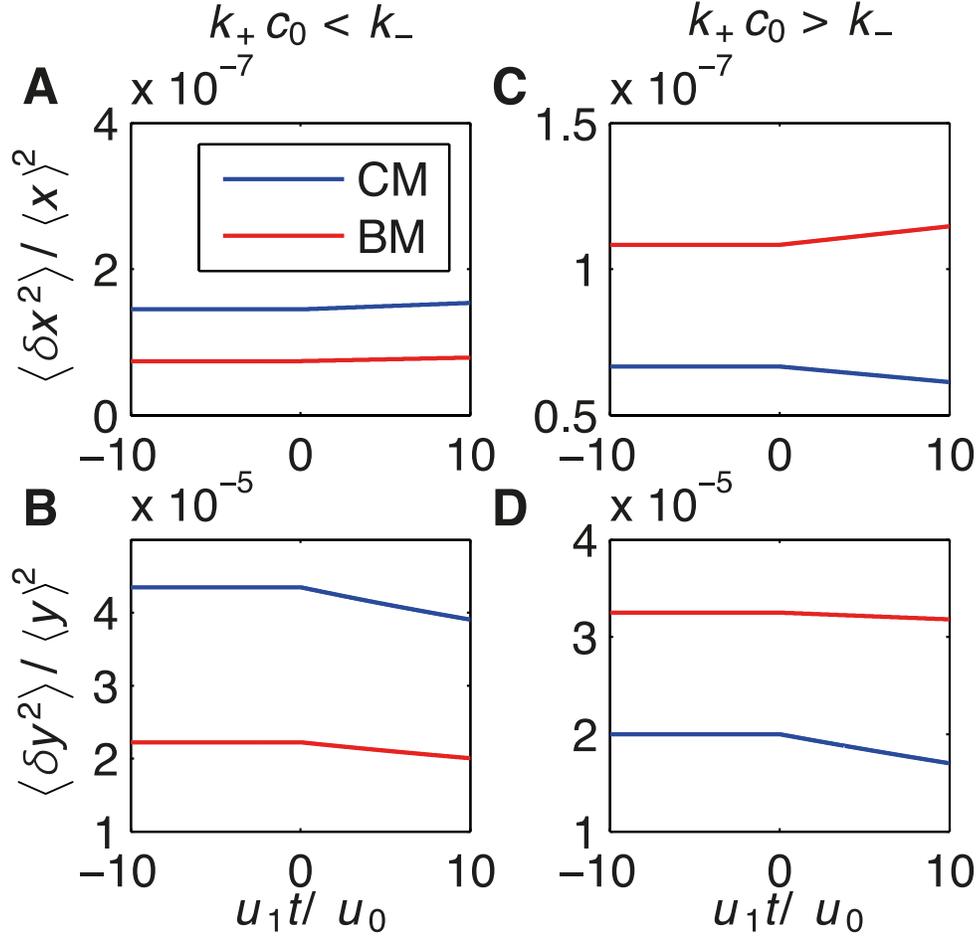} 
  \end{center}
  \caption{\textbf{Two regimes in incoherent feedforward loop based on the small-noise approximation.} Output noise, \textit{i.e.} relative variance of $x$ (top) and $y$ (bottom), as function of the non-dimensional ramp time $u_1 t / u_0$ for $k_+ c_0 < k_-$ \textit{i.e.} $\left< \tau_b \right> < \left< \tau_u \right>$ (left) and $k_+ c_0 > k_-$ \textit{i.e.} $\left< \tau_b \right> > \left< \tau_u \right>$ (right). CM and BM are shown by blue and red lines respectively. (A,B) BM is more accurate than AM for $k_+ c_0=10^{7} s^{-1}$ and $k_-=6.7 \times 10^{7} s^{-1}$. (C,D) CM is more accurate then BM for $k_+ c_0=10^{7} s^{-1}$ and $k_-=6.7\times 10^{6} s^{-1}$. Remaining parameters: $k_+ c_1= 10^5 s^{-2}$, $k_x=5 s^{-1}$ and $k_y=10 s^{-1}$.}
  \label{fig6}
\end{figure}

To analytically solve Eqs. \eqref{eq:incoherentFLx} and \eqref{eq:incoherentFLy} for the incoherent feedforward loop, and Eqs. \eqref{eq:integralFCx} and \eqref{eq:integralFCy} for the integral feedback loop, we linearize these equations within the small-noise approximation, and assume that we are in the fast-switching regime. This allows us to find analytic solutions in a particular time window and under certain conditions which we define in S1 Text. 
Specifically, for the incoherent feedforward loop in the small-ramp regime, the average values of $\left< u(t) \right>$, $\left< x(t) \right>$ and $\left< y(t) \right>$ are determined by the differential equations Eqs. (\ref{eq:incoherentFLx},\ref{eq:incoherentFLy}). Although there are no steady states for ramps, $\left< x(t) \right>$ and $\left< y(t) \right>$ show time-dependent stable solutions
%
%
\begin{subequations}
\begin{align}
\label{eq:Xstable}
\left< x(t) \right> &= e^{\frac{b u_1}{k_y}}, \\
\label{eq:Ystable}
\left< y(t) \right> &= \frac{u_0}{k_y} - \frac{u_1}{k_y^2} + \frac{u_1 t}{k_y}.
\end{align}
\end{subequations}
Introducing $x=\left< x\right> + \delta x$ and $y=\left< y\right> + \delta y$ into Eqs. \eqref{eq:incoherentFLx} and \eqref{eq:incoherentFLy} with subsequent linearization the variance of the target-protein copy numbers can be derived (see Materials and Methods). 
To first order in small-ramp parameters the variances of $x$ for both types of modulation are
\begin{subequations}
\begin{align}
\label{eq:varXsx}
\left< \left(\delta x(t)\right)^2 \right>_{\text{CM}} = & \ \Delta \left[ g_\text{CM} u_0 - \frac{1 - 2 c_0 \kp / k_-}{k_x + k_y}u_1 + g_\text{CM} (1 - 2 k_+ c_0 / k_-) u_1 t   \right], \\
\label{eq:varYsy}
\left< \left(\delta x(t)\right)^2 \right>_{\text{BM}} = & \ \Delta  \left[ g^*_{\text{BM}} u_0 - \frac{1 -2 c_0 \kp / k_- + 3 c_0^2 \kp^2 / k_-^2}{2 k_y} u_1 + \left( 1 - 2 \kp c_0 / k_-+ 3 \kp^2 c_0^2 / k_-^2 \right) u_1 t   \right] ,
\end{align}
\end{subequations}
where $\Delta=\frac{b^2 k_x^2 e^{\frac{2 b u_1}{k_y}}}{2 (k_x+k_y)(1+\kp c_0 / k_-)^2}$, and $g^{\phantom *}_\text{CM}$ and $g^*_\text{BM}$ are parameters discussed in Eqs. \eqref{eq:g} and \eqref{eq:defg*}. The corresponding results for species $y$ are provided in Eqs. (S56) and (S58), and plots for species $x$ and $y$ are shown in Fig. \ref{fig6}B,D.

Consistent with the master equation, these results show again two regimes: ramp sensing is more accurate for BM if $k_+ c_0 < k_-$, while CM is more accurate otherwise. For a constant environment (zeroth-order with $c_1=u_1=0$) the regime is largely determined by the factor $g$. If $\kp c_0 < k_-$, $g_{\text{BM}}=1+\left<(\delta \tau_b)^2\right>/\left< \delta (\tau_u)^2 \right> < 2$ (see Eq. \eqref{eq:g}), and BM is more accurate than CM with $g_{\text{CM}}=1+\left<(\delta \tau_b)^2\right>/\left< \tau_b \right>^2=2$ (Fig. \ref{fig6}A,B). 
This is because the variability of the bound intervals $\left< (\delta \tau_b)^2\right>$ can be eliminated in BM (but not in CM), and the unbound intervals are well approximated by the duration between bursts ($\tau_{\text{bursts}}$ in Fig. 3). For $k_+ c_0 \ll k_-$, BM effectively implements MLE. In contrast, CM is more accurate for $k_+ c_0 > k_-$, where $g_{\text{CM}}=2$ and $g_{\text{BM}}>2$ (Fig. \ref{fig6}C,D). This is because BM contains no information on unbound time intervals, while CM still contains some information through the probability of being bound ($p_b$ in Fig. 3B). These results also apply to ramp sensing since the accuracy of the downstream proteins (decoding) relates again to the factor $g$ and hence to the ratio between the bound and unbound time intervals.
The integral feedback loop in Eqs. \eqref{eq:integralFCx} and \eqref{eq:integralFCy} shows very similar behavior (provided in S1 Text). The validity of our analytical results are confirmed by simulations of the stochastic differential equation for both pathways in \ref{figS1}. and \ref{figS2}.\\

\subsection*{AM is more accurate than FM for multiple receptors/ion channels}

To address the question of whether AM or FM is more accurate in encoding and decoding, we consider a straightforward generalization to multiple receptors (or ion channels) (see S1 Text and \ref{fig5}. for details). AM can be obtained by considering unsynchronized CM receptors. In contrast, the experimentally observed sporadic bursts of nuclear translocation \cite{OShea,MW} and hence FM might be explained by synchronized receptors that individually operate with BM.

For $N$ unsynchronized (\textit{us}) receptors, the resulting average and variance of the signaling rate are $\left< u(t) \right>_N^{us} = N \left< u(t) \right>_1$ and $\left< \delta u(t) \delta u(t') \right>_N^{us} = N \left< \delta u(t) \delta u(t') \right>_1$ in terms of the single-receptor quantities. Consequently, the relative variance, given by the variance divided by the average-squared, scales with the inverse of the number of receptors ($N$). On the other hand, for $N$ synchronized (\textit{s}) receptors, the average and variance of the signaling rate are given respectively by $\left< u(t) \right>_N^{s} = N \left< u(t) \right>_1$ and $\left< \delta u(t) \delta u(t') \right>^s_N = N^2 \left< \delta u(t) \delta u(t') \right>_1$. 
The relative variance is now independent of $N$. Hence, unsynchronized receptors (AM) have a reduction of noise by a factor $N$ compared to synchronized receptors (FM). 

For slow dynamics, or fast dynamics with $k_+ c > k_-$, CM is generally more accurate than BM (at least for $\zeta>1$), and with $N$ receptors, AM is more accurate than FM by an even larger margin. In contrast, for fast dynamics with $k_+ c < k_-$, BM is more accurate than CM by at most a factor of 2 (Eq. \eqref{eq:g}). But since AM is $N$ times more accurate than CM, AM becomes more accurate for encoding than FM for more than two receptors. Since our results from the previous sections show that larger signaling noise leads to larger output noise, the same rule emerges for decoding.

From a physical point of view, how can receptors act in a synchronized fashion? Receptors may be coupled by adaptor proteins or elastic membrane deformations, allowing them to act cooperatively \cite{Endres_2009BiophysJ,Wiggins_05BiopJ}.
In conclusion, while for fast dynamics (small-noise approximation) BM can be more accurate than CM up to a factor of two, two receptors/ion channels are sufficient for AM to become more accurate than FM. Since cells have thousands of receptors and ion channels, AM becomes the most accurate modulation scheme. \\


\section*{Discussion}

Cellular responses to extra-cellular stimuli involve both encoding the external stimuli by internal signals (which is normally fast) and subsequently decoding via the regulation of protein levels (which is normally much slower). The internal representation of the external signal falls into two broad categories: continuous/amplitude modulation (CM/AM), where bound receptors continually signal and the internal concentration itself encodes the external signal, and bursty/frequency modulation (BM/FM), where receptors only signal when first bound and the signal is encoded in the frequency of peaks. Here, we compared the output noise for both types of modulation in the presence of a constant and a linearly increasing (in time) external concentration. Besides considering a linear pathway, we compared two nonlinear network motifs: the incoherent feedforward loop and the integral feedback loop. These loops are ubiquitous in biological systems. For example, the incoherent feedforward loop is found in chemotactic adaptation of eukaryotes \cite{Takeda_2012} and transcription networks in bacteria \cite{Shen_Alon}, and the integral feedback loop is found in chemotactic adaptation of bacteria \cite{Doyle_onIFC, Alon_99} and in eukaryotic olfactory and phototransduction pathways \cite{Giovanna_13}.

We found that, for a single receptor or ion channel, BM can be more accurate than CM for fast dynamics.  This situation can occur when the average duration of the active \textit{on} state is shorter than the average duration of the inactive \textit{off} state (Figs. \ref{fig8} and \ref{fig6}). In this case, BM effectively implements maximum-likelihood estimation, the most accurate mechanism of sensing \cite{EndWing}. 
If instead more time is spent in the \textit{on} state, then CM is generally more accurate (except when the burst size is minimal, i.e one). 
The reason behind this effect, which we analytically prove within the small-noise approximation, is that CM has information about both the \textit{on} and \textit{off} states, whereas BM only knows when a switch from \textit{off} to \textit{on} occurs. As such, CM effectively implements Berg and Purcell's classic result of estimating ligand concentration by time averaging \cite{Berg_Purcell} (see also Discussion in \cite{EndWing}). In addition, we found that for slow dynamics CM is always more accurate than BM, independent of whether more time is spent in the \textit{on} or \textit{off} states, due to increased burst sizes (Fig. \ref{fig8}). Taken together our results suggest that BM should be more common in signaling pathways than in gene regulation.

The generalization to multiple receptors/ion channels allows AM and FM to be compared. AM, which arises from unsynchronized CM receptors, has a reduced relative noise due to spatial averaging, while the relative noise in FM from synchronized BM receptors remains identical to the single-receptor result. (Note the observed nuclear bursts of approximately constant amplitude and duration support our FM mechanism \cite{Elowitz_08,OShea}.)  As a result, AM is always more accurate than FM for more than two receptors (\ref{fig5}.). Since cells have tens of thousands of receptors and ion channels, this implies that the reason that FM is sometimes observed in real systems must have a different origin. At least three possibilities present themselves. 
Firstly, FM can help to coordinate gene expression \cite{Elowitz_08, Eldar_Elowitz_10}, which is particularly useful when hundreds of genes are controlled by a single transcription factor, such as during stress response \cite{diffT,diffT2,diffT3}. Secondly, FM can enhance co-localization of proteins inside the nucleus, providing another way to improve coordination of gene expression \cite{Kang_PLoSCB}. Thirdly, as with oscillatory signals, bursts can be used to activate transcription by threshold crossing \cite{Dolmetsch_1998} while avoiding desensitization \cite{Li_Goldbeter}. This may then push the cell to differentiate into a new state (such as under starvation to initiate competence) \cite{Ferrell_2012,Purvis_Science_12}. It is also worth noting that by using seemingly redundant isoforms (such as NFAT1 and NFAT4 during an immune response), AM and FM can be combined to enhance temporal information processing \cite{Yissachar2013}.

While providing intuitive insights, it is clear that our models are highly oversimplified versions of signaling and gene regulation in actual cells. One of the main reasons for this is that we used idealized delta-functions as pulses in BM (and hence in FM). However, for example, in the calcium stress-response pathway in \textit{Saccharomyces cerevisiae} (Fig. \ref{fig3}A) nuclear bursts of Crz1p are on average two minutes long (Fig. 1D, inset). Most likely cytoplasmic calcium spikes determine the nuclear bursts (Elowitz, personal communication), but since the mechanism of calcium spiking remains poorly understood, such bursts are difficult to model. A further limitation of our models is that bursts only relate to translocation, whereas additional bursts may occur further downstream during transcription \cite{Muramoto_2012PNAS} (e.g. due to promoter switching \cite{Wolde_12}) and translation \cite{Thattai2001}. Future models may need to include these details.

Our models suggest further experimental investigation in multiple areas. Firstly, the distribution of burst duration affects factor $g$ (Eq. \eqref{eq:g}), 
so that $g = 2$ in equilibrium for a single-step process and potentially $g<2$ for an irreversible binding cycle dominated by energy dissipation \cite{MW,Lang_14PRL}. These irreversible cycles are present in some ligand-gated ion channels, such as 
 the cystic fibrosis transmembrane conductance regulator (CFTR) channels and N-Methyl-D-aspartate (NMDA) receptors. These exhibit peaked opening distributions, which can be interpreted as evidence of broken reversibility and energy consumption \cite{Csanady_2010,Schneggenburger_1997}. Such cases and their possible connection with accuracy need further investigation. In fact, most cellular processes rely heavily on energy consumption, including nuclear shuttling and chromosome remodeling, limiting the applicability of our equilibrium CM-receptor model. Secondly, coordination of gene expression during stress or cell-fate decisions might be another reason for implementing FM rather than AM. More quantitative experiments are needed to better understand this mechanism. Thirdly, closer inspection of $\text{Ca}^{2+}$-independent transcription factors (as well as $\text{Ca}^{2+}$-dependent co-regulated genes) are warranted in order to verify coordination of multiple genes \cite{Elowitz_08}. Finally, to see if bursts help jump start new cellular programs (\textit{i.e.} transition into a new ``attractor''), global changes in gene regulation can be monitored.

A general understanding of FM may help prevent developmental defects and human diseases. Indeed, several biomedically relevant transcription factors, such as $\text{NF-}\kappa$B, p53, NFAT and ERK, show oscillatory pulsing or random bursting \cite{Burst,Levine2013, Purvis_Science_12, Batchelor_2011, Albeck2013, Tay2010,Ashall_Science}. In fact, the destabilization of regulatory circuits can underlie human diseases: studies suggest that the coordination of gene expression could be critical in maintaining the proper functioning of key nodes in such circuits. For example, the NFATc circuit is cooperatively destabilized by a 1.5-fold increase in the DSCR1 and DYRK1A genes, which reduce NFATc activity leading to characteristics of Down's syndrome \cite{Arron_2006Nature,Burst}. However, ERK pulses are regulated by both AM and FM with the same dose dependence, and it remains unclear how they affect cell proliferation and the relevance to cancer \cite{Albeck2013}.

Broadly speaking, temporal ordering (regularity or periodicity) serves at least two roles in living systems \cite{Anderson72}: extraction of energy from the environment and handling of information. While the first role is well studied in terms of molecular motors at the single-molecule level, the second role is intellectually more difficult to understand as it requires a broader, more global understanding of cells. We believe that future work that combines single-cell experiments with ideas of collective behavior and engineering principles is most likely to be successful.


\section*{Materials and Methods}

\subsection*{Master-equation model for concentration sensing}

The master equations for continuous modulation (CM), Eqs. \eqref{eq:masterAMon} and \eqref{eq:masterAMoff}, can be solved at steady state using generating functions. In particular, we derive the first three moments of the probability distribution using the general model in \cite{Metha_Schwab}. When the system is in the \textit{on/off} state, the production rate of species $x$ is $\alpha_\textit{on/off}$. The degradation rate $\gamma$ is independent of the state of the system. The probability distribution of $n$ target proteins at time $t$ is then described by
  \begin{align}
  \label{eqmet:masterMS}
  \frac{dp_s(n,t)}{dt} =& \, \gamma (n+1) p_s(n+1,t) + \alpha_s p_s(n-1,t) + k_{\bar{s}} p_{\bar{s}}(n,t) - (\gamma n + \alpha_s + k_s) p_s(n,t) ,
\end{align}
where $\bar{s}=\text{off} \ (\text{on})$ when $s=\text{on} \ (\text{off})$. By defining the generating functions
\begin{align}
  \label{eqmet:Gfun_MS}
  G_s(z)=\sum_{n=0}^{\infty} p_s(n) z^n,
\end{align}
and using Eq. \eqref{eqmet:masterMS}, a solution for $G_s(z)$ can be found, which then readily gives the moments of $p(n,t)$. In particular, the variance and skewness are given by
\begin{align}
  \left< \delta n ^2 \right> = & \sum_s \left( \partial_z z \partial_z G_s(z) \right)\big|_{z=1} - \left< n \right> ^2, \\
  \left< n^3 \right> = & \sum_s \left[ \partial_z z \partial_z z \partial_z G_s(z) \right]\Big|_{z=1}.
\end{align}
Full details are given in S1 Text.

In order to solve the master equation for bursty modulation (BM), Eqs. \eqref{eq:masterFMon} and \eqref{eq:masterFMoff}, we use the Gillespie algorithm \cite{Gillespie}. If the system is in the \textit{on} state with $n$ proteins at time $t$, it can either switch to the \textit{off} state with transition rate given by $k_-/(k_-+\gamma n)$ or else remain in the \textit{on} state and lose a protein by degradation. If instead the system is in the \textit{off} state with $n$ proteins at time $t$, it can either switch to the \textit{on} state with switching rate $k_+ c_0/(\kp c_0+\gamma n)$ and, via a burst, increase its number of proteins to $n+\zeta$, or again remain in the same state and loose a protein by degradation. The time step between reactions, $\delta t$, is chosen from an exponential probability distribution $\lambda e^{- \lambda \delta t}$, with $\lambda$ equal to the total rate that at least one reaction occurs.

\subsection*{ODE models for ramp sensing}

The following method applies to both the incoherent feedforward and the integral feedback loop. To solve the ordinary differential equations (\ref{eq:incoherentFLx}-\ref{eq:integralFCy}) we linearize around stable solutions, $x(t)=\left< x(t) \right> +\delta x$ and $y(t)=\left< y(t) \right> +\delta y$, and assume that small $\delta u$ leads to small $\delta x$ and $\delta y$. Note that when sensing a gradually changing ramp, $\left< x(t) \right>$ and $\left< y(t) \right>$ are not steady states. Defining $X=[x(t)\>y(t)]^T$ we can rewrite these equations as
\begin{equation} \label{eq:Xeq}
  \frac{dX(t)}{dt} + M X(t) = \begin{bmatrix} w \ \delta u \\  z \ \delta u \end{bmatrix},
\end{equation}
where the matrix $M$ and the constants $w$ and $z$ are defined in S1 Text. Analytic solutions are only available when $M$ is time-independent. As shown in S1 Text, Eq. \eqref{eq:Xeq} can be solved and written as an integral, which can then be evaluated with, for example, Wolfram Mathematica 8.


\section*{Acknowledgments}
We thank Guido Tiana for helpful discussions, and Michael Elowitz for providing details on his data. GM and RGE acknowledge funding from ERC Starting-Grant N. 280492-PPHPI. GA was supported by Leverhulme-Trust grant N. RPG-181 and DMR was supported by BBSRC grant N. P34460.




\newpage

\renewcommand{\figurename}{}

\renewcommand\thefigure{S\arabic{figure} Fig}    
\renewcommand\thetable{S\arabic{table}}    
\setcounter{table}{0}
\setcounter{figure}{0}

\section*{Supporting Information Legends}

\paragraph{ S1 Text.} Details of analytical calculations. 

\begin{figure}[!ht]
  \begin{center}
    \end{center}
 \caption{\textbf{First three moments of the protein distribution in concentration sensing from the master equation.} Averages (A,B), variance (C,D), and skewness (E,F) as a function of the frequency of binding events, $f=k_+ c_0/(1+k_+ c_0/k_-)$. (Insets) Magnification of small-noise approximation region (fast switching). Analytical results for CM (blue) and numerical results for BM (red) and intermediate modulation IM (green) as function of the frequency of binding events (logarithmic scale). Note that this figure is similar to Fig. \ref{fig8} in main text with the addition of IM. Two regimes are shown: $k_-=10 \ k_+ c_0$ ($\alpha=100 s^{-1}, \gamma=1 s^{-1}$, $\zeta \text{ from } 1000 \text{ to } 1$) (left column) and $k_-= 0.1 \ k_+ c_0$ ($\alpha=10 s^{-1}, \gamma=1 s^{-1}$, $\zeta \text{ from } 1000 \text{ to } 1$) (right column). Averages from CM, BM and IM are constrained to be equal, \textit{i.e.} $\zeta$ (BM) $=\alpha k_-^{-1}$ (CM) $=\alpha' \tau_b$ (IM). Variances of CM, BM and IM exhibit two different regimes for fast switching: for $k_+ c_0 < k_-$ BM is the most accurate mechanism and CM the worst (inset in C) while for $k_+ c_0 > k_-$ CM is generally the most accurate (except for $\zeta=1$) and IM the worst (inset in D). Third moments show that, for large noise, the probability distributions become asymmetric.}
  \label{fig5like}
\end{figure}

\begin{figure}[!ht]
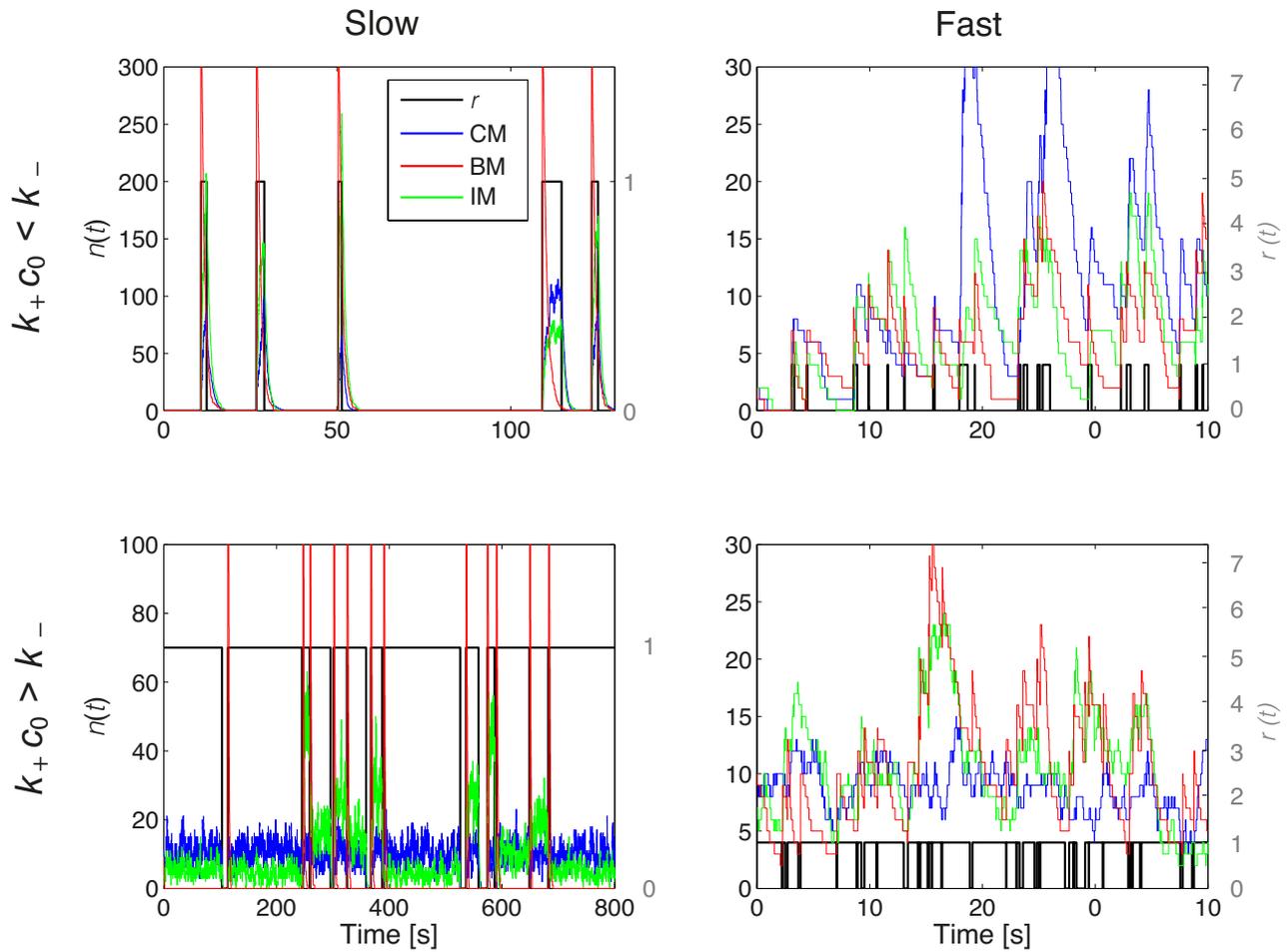

  \begin{center}
    \end{center}
 \caption{\textbf{Examples of time traces of receptor activity and protein copy numbers for different regimes.} (Top) Regime $k_+ c_0 < k_-$ with $k_+ c=0.1 \ k_-$ ($\alpha=100 s^{-1}, \gamma=1 s^{-1}$). (Bottom) Regime $k_+ c_0 > k_-$ with $k_+ c=10 \ k_-$ ($\alpha=10 s^{-1}, \gamma=1 s^{-1}$). (Left) Slow switching with $\zeta=400$. (Right) Fast switching with $\zeta=7$. Receptor activity $r$ and protein copy numbers $n(t)$ for CM, BM and IM are shown in black, blue, red and green, respectively.}
  \label{figSItraj1}
\end{figure}

\begin{figure}[!ht]
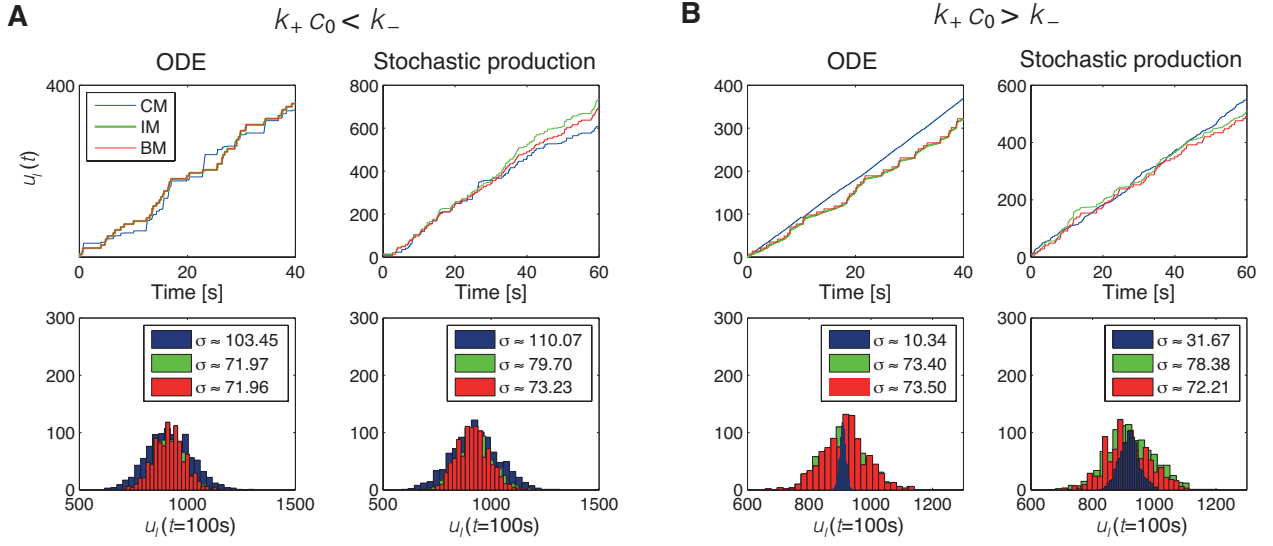

  \begin{center}
    \end{center}
 \caption{\textbf{Investigating accuracy based on accumulative signaling (without protein production and degradation).} (A) Regime $k_+ c_0 < k_-$ with $k_+ c=0.1 \ k_-$ ($\alpha=100 s^{-1}, \gamma=1 s^{-1}$ and $\zeta=7$). (Left) ODE model. (Right) Stochastic protein production during $\tau_b$ in CM and IM. (Top) Examples of time traces. (Bottom) Histograms of number of proteins produced after 100s with standard deviation in legend based on 1000 simulations. (B) Analogous to (A) but for regime $k_+ c > k_-$ with $k_+ c=10 \ k_-$ ($\alpha=100 s^{-1}, \gamma=1 s^{-1}$ and $\zeta=7$). CM, BM and IM are shown in blue, red and green, respectively.}
  \label{figSIstoc1}
\end{figure}

\begin{figure}[h]
    \caption{\textbf{Incoherent feedforward loop: Comparison of analytical results with simulations of the stochastic differential equations.} (A) Averages of signaling rate $u$ (left), species $y$ from Eq. (S42) (middle) and species $x$ from (S41) (right) as a function of time. Analytic solutions Eqs. (S32), (S43) and (12) are shown for BM in red, while a (time averaged) time-trace from a stochastic simulation using the Euler method is shown in orange (CM is almost identical and hence is not shown). (B) Corresponding variances as a function of time for $k_+ c_0 > k_-$ ($k_-=6.7 \times10^{5} s^{-1}$, $k_+ c_0=10^{6} s^{-1}$). Analytic results are shown in blue for CM and in red for BM; average over time ($1s$) from numerical simulations are shown in light blue for CM and in orange for BM. (C) Corresponding variances as a function of time for $k_+ c_0 < k_-$  ($k_-=6.7 \times10^{6} s^{-1}$, $k_+ c_0=10^{6} s^{-1}$). Colors same as in (B). Remaining parameters: $k_+ c_1= 10^4 s^{-2}$, $k_x=10 s^{-1}$ and $k_y=50 s^{-1}$.
  \label{figS1}}
\end{figure}

\begin{figure}[h]
  \caption{\textbf{Integral feedback loop: Comparison of analytical results with simulations of the stochastic differential equations.}  (A) Averages of signaling rate $u$ (left), species $y$ from Eq. (S60) (middle) and species $x$ from (S59) (right) as a function of time. Analytic solutions Eqs. (S32), (S66) and (S65) are shown for BM in red, while a (time averaged) time-trace from a stochastic simulation using the Euler method is shown in orange (CM is almost identical and hence is not shown).  (B) Corresponding variances as a function of time for $k_+ c_0 > k_-$ ($k_-=6.7 \times10^{5} s^{-1}$, $k_+ c_0=10^{6} s^{-1}$). Analytic results are shown in blue for CM and in red for BM; numerical simulations are shown in light blue for CM and in orange for BM. (C) Corresponding variances as a function of time for $k_+ c_0 < k_-$  ($k_-=6.7 \times10^{6} s^{-1}$, $k_+ c_0=10^{6} s^{-1}$). Colors same as in (B). Remaining parameters: $k_+ c_1= 10^4 s^{-2}$, $k_x=10 s^{-1}$ and $k_y=50 s^{-1}$.
  \label{figS2}}
\end{figure}

\begin{figure}[!ht]
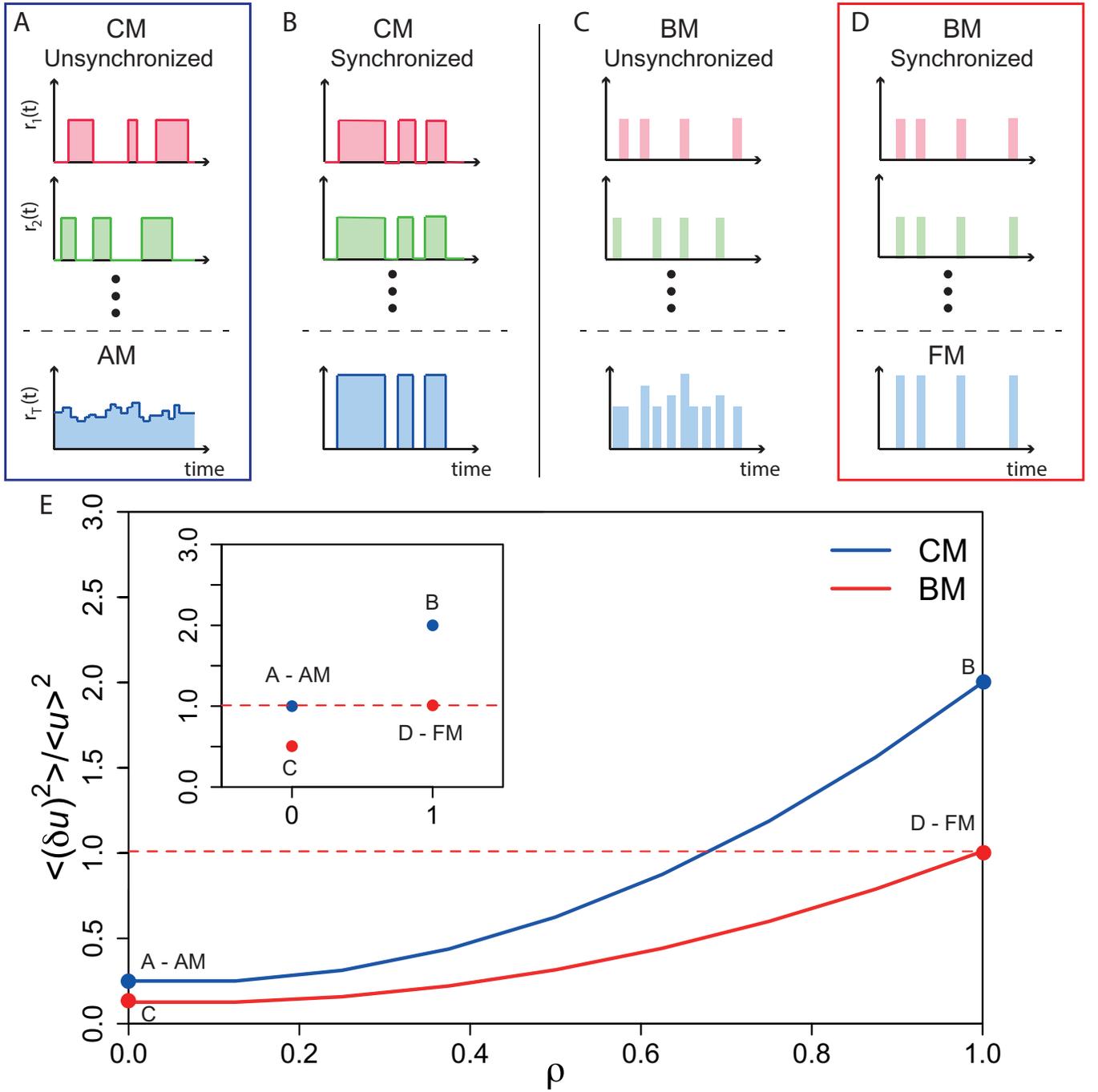

  \begin{center}
    \end{center}
 \caption{\textbf{From CM (BM) to AM (FM) for multiple receptors/ion channels.} (A-D) Schematic of receptor activity in time. (A) AM emerges from $N$ unsynchronized receptors or ion channels in CM mode. (B) $N$ synchronized CM receptors lead to a hybrid mechanism with information encoded in the frequency of broad bursts of variable duration. (C) $N$ unsynchronized BM receptors provide a dense series of bursts. For large $N$, bursts may start overlapping, leading to variable amplitudes. (D) FM emerges from $N$ synchronized receptors in BM mode. (E) Relative variance for a system of 8 receptors with $\rho N$ synchronized and $(1-\rho) N$ unsynchronized receptors, plotted for fast dynamics in the $k_+ c < k_-$ regime (CM in blue and BM in red). Letters refer to panel labels (A-D). Dotted red line indicates uncertainty from FM for comparison. (Inset) Same for a system of two receptors only.}
  \label{fig5}
\end{figure}

\end{document}